\documentstyle[twocolumn,prl,aps,epsfig]{revtex}
\begin{document}
\draft
\twocolumn[\hsize\textwidth\columnwidth\hsize\csname@twocolumnfalse%
\endcsname

\title{Suppression of tunneling into multi-wall carbon nanotubes}

\author{A. Bachtold$^*$, M. de Jonge, K. Grove-Rasmussen, P.L. McEuen}

\address{Department of Physics, University of California at Berkeley and 
Materials Sciences Division, Lawrence Berkeley National Laboratory, 
Berkeley, CA 94720, USA
}

\author{M. Buitelaar, C. Sch\"onenberger}

\address{Institut f\"ur Physik, Universit\"at Basel, 
Klingelbergstr. 82, CH-4056 Basel, Switzerland
 }

\date{December 5, 2000} \maketitle

\begin{abstract}
We have studied tunneling of electrons into multi-wall 
carbon nanotubes. Nanotube/electrode interfaces with low 
transparency as well as nanotube/nanotube junctions 
created with atomic force microscope manipulation have 
been used. The tunneling conductance goes to zero as 
the temperature and bias are reduced, and the functional 
form is consistent with a power law suppression of tunneling 
as a function of energy. The exponent depends upon sample 
geometry. The relationship between these results and theories 
for tunneling into ballistic and disordered metals is discussed. 
\end{abstract}
 ]

Tunneling into one-dimensional or quasi-one dimensional conductor 
is in general a complex process. Not only must the electron 
tunnel through the classically forbidden region of the tunnel 
barrier, room must be made in the conductor for the excess 
charge induced by the tunneling. This process is typically 
described by coupling the electron to a set of harmonic oscillators 
with a linear dispersion, resulting in power-law behavior of 
the conductance as a function of the tunneling energy of the 
electron. The values of the power law exponent are determined by 
the tunneling geometry and the properties of the oscillators. 
Many examples of this phenomenon have been discussed. For example, 
the 1D electromagnetic modes of microscopic resistors can be modeled 
in this manner\cite{ingold}, as can the collective excitations 
(plasmons) of the interacting 1D electron gas\cite{fisher}.
   
Carbon nanotubes are emerging as an excellent system for the 
investigation of electronic transport in 1D.  Two different 
classes of nanotubes exist: small diameter ($\approx$1nm) single-wall 
nanotubes (SWNTs) and large diameter ($\approx$10nm) multi-wall 
nanotubes (MWNTs). Metallic SWNTs are characterized by two 1D 
channels, and they can have a very long mean free 
path\cite{white,tans,bachtold1}: $l > 1 \mu m$.  As such, they represent a 
nearly perfect 1D system. It is well 
known theoretically that in 1D, 
transport is strongly affected by electron-electron 
interactions, producing a system called a 
Luttinger-liquid\cite{fisher,egger1} (LL) whose low 
energy states are collective in nature. The tunneling 
density of states of a LL diminishes as a 
power law on the energy of the tunneling electron. 
Experiments\cite{bockrath,yao} on SWNTs have shown that 
the tunneling conductance vanishes as a power law in 
temperature and in bias voltage. The value of the exponent 
has been found to depend on the tunneling geometry, 
i.e. whether the electron tunnels into the end or the bulk 
of the tube or between the ends of two tubes, in agreement 
with theory. 

Various experiments on the transport properties of MWNTs 
have been performed. Experiments with low ohmic contacts 
attached to MWNTs show interference effects, like Aharonov-Bohm 
oscillations\cite{bachtold2,schonenberger}, 
weak localization\cite{schonenberger,langer}
and universal conductance fluctuations\cite{schonenberger,langer}. 
These have been used to infer the mean free path, yielding 
$l \sim 5-150 nm$\cite{schonenberger}. They also indicate 
that the current may be predominantly carried in the outer shell 
of the nanotube. Similarly, transport\cite{schonenberger} and scanned 
probe experiments\cite{bachtold1,dai} indicate a typical resistance 
per unit length of 5-10 kOhm/$\mu$m.

We present here measurements of the tunneling 
conductance of MWNTs. In one geometry, metallic 
contacts to the tube with high resistance are employed. 
In a second geometry, nanotube/nanotube junctions are created 
by manipulation with an atomic force microscope. The tunneling 
conductance of the nanotube/electrode interface is measured as 
a function of temperature and bias voltage. These measurements 
show that the tunneling density of states diminishes as 
power laws and that the exponent depends upon on the geometry. 

The MWNTs were synthesized by arc-discharge evaporation and 
deposited from a dispersion in chloroform onto an oxidized Si wafer. 
Nanotubes with diameter ranging from 8 to $17nm$ are 
selected and located using a scanning 
electron microscope or an AFM.  For devices of 
the first type, gold contacts to the tube are 
then created using e-beam lithography. 
This procedure typically leads to contacts with low resistance 
($\approx1k\Omega$), but it also occasionally produces 
highly resistive contacts ($ > $10k$\Omega$). The microscopic 
origin of the high resistance is not known, but 
a variety of causes are possible. For example, a thin 
insulating layer of organic contamination may be on the tube. 
Here, however, we will merely exploit these accidental tunnel 
barriers to probe the electronic properties of the MWNT.

Inset of Fig.~1 shows the schematic of a device consisting of a $d=17nm$ 
diameter MWNT contacted to 3 electrodes. The 4.5k$\Omega$ 
resistance measured between the two outer electrodes corresponds 
to the typical resistance for a MWNT contacted with electrodes 
that are separated by $700nm$. This indicates that the contact 
resistances of the two outer electrodes are low and 
that the nanotube connecting them is electrically continuous. 
However, the resistance measured from the inner electrode to 
either of the outer electrodes is much higher, 140k$\Omega$. 
The metal/nanotube interface at the inner electrode has 
thus a low transparency and serves as a tunneling barrier into the nanotube. 
We therefore use this contact to measure the tunneling 
density of states of the tube.

Fig.~1 shows a series measurements of the tunneling conductance 
versus bias for different magnetic fields $B$ applied parallel 
to the tube. The $dI/dV$ spectra are highly structured. 
We first note that, at small energies (0-10mV) 
the spectra display a strongly suppressed conductance. 
This suppression is independent of $B$. This suppression 
is the main subject of this paper, but we will 
first address the complex features at higher energy.
The high bias features are quasi-periodic, 
with an average spacing between the maxima of
$\approx$25mV. In addition, these features 
evolve with increasing magnetic field.  

The behavior of the high-bias features is consistent 
with expectations for the effects of subbands on 
tunneling into a quasi-1D conductor. This typical spacing of 
the peaks (25 mV) is in reasonable agreement with 
the theoretical separation expected for van Hove peaks 
for a 17nm diameter nanotube.  The average separation 
is predicted to be $\hbar v_{F}/d$ = 29meV, where $v_{F}$ = $8~10^{5}$ m/s 
is the Fermi velocity. We take $d$ as the diameter of the 
outermost cylinder, because the current has been shown 
to be carried predominantly by the outermost cylinder. 
The emergence of (broadened) van Hove peaks in the 
tunneling density of states demonstrates that the elastic 
length is at least of the same order as the circumference of the tube. 
On the other hand, the substantial peak broadening and the lack of 
accurate periodicity of the peaks indicate the disorder is playing an important role. 

The identification\cite{note1} of these peaks as broadened van Hove 
peaks\cite{wildoer} is further supported by their behavior in $B$. 
The peaks move up and down in parallel field, 
as expected from the Aharonov-Bohm effect\cite{ajiki}. The magnetic 
flux through the nanotube gives 
rise to a phase that shifts the van Hove 
peaks\cite{note2} in energy. This results in a bandstructure 
that is periodic in magnetic flux with the fundamental period 
$h/e$ which would correspond to a field of 17.3T. 
However, the measured spectra in Fig.~1 do not appear to be periodic 
in field. The reason is not clear yet, but may be attributed to 
scattering, to the nanotube deformation or to the interaction 
between adjacent cylinders.  

We now turn to the low bias suppression of the conductance. 
It is independent of $B$, which suggests a different origin 
than the peaks discussed above. In addition, its dependence 
on $V$ and $T$ is very different, as we now discuss.  

Fig.~2a shows the tunneling conductance of a second 
sample as a function of $V$ at a series of $T$.  The MWNT is 12nm 
in diameter and is attached to 3 contacts; 
the two outer electrodes are low resistance and 
the inner one has a resistance of 42k$\Omega$ at 300K. 
Again, this middle electrode is used to 
measure the tunneling conductance. 

A conductance dip centered at zero energy is again observed. 
As $T$ is decreased down to 350mK, the amplitude of the dip increases. 
In Fig.~2b the zero bias conductance is 
plotted as a function of $T$ in a double logarithmic plot. 
The straight-line behavior indicates that measured data 
is well described by a power law behavior $G = T^{\alpha}$ with exponent 
$\alpha$ = 0.36. For bias voltages larger than $k_{B}T/e$, the voltage 
dependence of the conductance can also be described by a power law 
with an exponent, which is again equal to 0.36. This can be seen in 
Fig.~2c, which shows the symmetrical part of the conductance divided 
by $T^{\alpha}$ as a function of the voltage divided by $k_{B}T/e$. 
This figure shows also that the scaled curves for different $T$ 
collapse well into a single universal curve\cite{bockrath}. 
A similar power-law scaling in $T$ and $V$ is found in 11 different 
samples with exponents $\alpha$ ranging from 0.24 to 0.37.

To explore this suppression of the tunneling conductance further, 
we created devices composed of two MWNTs arranged in different geometries. 
An AFM tip has been used to move nanotubes\cite{hertel,postma} 
using the approach discussed in Ref~18. The end of one tube is 
pushed against either the end or the middle (bulk) 
of a second tube. Au contacts are attached to both tubes. 
Examples of end-bulk and end-end junctions are shown in Fig.~3a and b. 
The resistance of these junctions varies considerably 
from device to device, from immeasurably large to $\approx$100k$\Omega$. 
This indicates that the junction between two tubes serves as a tunnel barrier.

As with the metal/nanotube junctions above, reducing $T$ and $V$ causes 
the conductance to decrease significantly, extrapolating 
to zero at zero temperature. This suppression is significantly 
more dramatic than in the metal/nanotube junction devices.  Fig.~3c shows 
the $dI/dV$ as a function of $V$ in a double logarithmic scale at $T$ = 3K 
for a bulk-end and an end-end junction. 
For comparison, the tunneling conductance curve of a nanotube-electrode 
junction is also plotted. The curve show approximate power 
law behavior, but with different values for the exponent. We find $\alpha$ = 0.9 
and 1.24 for bulk-end and end-end junctions, respectively. 
These exponents are representative of the seven junctions that we have studied. 

Overall, our result for tunneling into MWNTs can be summarized in a 
simple rule. The conductance is given by $G \sim E^{\alpha}$, where $E$ 
is the excess energy of the tunneling electron, given by the larger of 
$eV$ or $k_{B}T$.  To a first approximation, the exponent $\alpha$ of the 
power law is equal to $\alpha = \alpha_{1} + \alpha_{2}$, where the terms 
$\alpha_{i}$ represent the properties of the conductor on 
each side of the junction. We find that $\alpha_{i}$ $\sim$ 0.3 for the tube bulk, 
$\alpha_{i}$ $\sim$ 0.6 for the tube end, and $\alpha_{i}$ $\sim$ 0 for the Au contact.

We now discuss the possible origins of this low energy suppression of 
the conductance. The first possbility is the single particle density of 
states (DOS) of the graphene, the 2D material from 
which nanotubes are made. The DOS is linearly dependent on 
energy. Indeed, some samples show a linearly increasing $dI/dV$ at high $V$ 
($ > $100 mV), that can be attributed to this. 
At low energies, however, the DOS should be significantly 
modified. First, the quantization of the energy states 
due to the transverse confinement should create a 
series of 1D subbands as discussed above. On energy scales 
much less than the subband spacing, the DOS should be relatively 
constant. This is however not observed.  In addition, the position 
in voltage of the minimum in $dI/dV$ would depend on gate 
voltage, since the gate voltage shifts the Fermi energy. 
Again, this is not observed. We therefore conclude that the 
anomaly is not related to the band structure of graphene.

Another possibility is tunneling anomalies associated with 
Coulomb interactions that were discussed in the introduction. 
These lead to power-law dependences: $dI/dV \sim V^{\alpha}$ and 
$G \sim T^{\alpha}$ at zero temperature and zero voltage, respectively. 
Theoretically,$\alpha$ depends on the 
junction geometry and the properties of the 
conductor. 
In one case the conductor is approximated as a 
classical resistor R\cite{delsing}, and the exponent is then 
given by: $\alpha = R/(h/2e^{2})$ where $R$ is the effective 
resistance seen by the tunneling electron. 
The second case applies for a ballistic Luttinger liquid. 
The exponents for tunneling into the end and bulk of the 
conductor are predicted to be $\alpha^{end}=(g^{-1}-1)/2$ 
and $\alpha^{bulk} = (g^{-1}+g-2)/4$ where $g$ 
depends on the interaction strength in the conductor. 
Both cases, a resistor and a LL, result in power-law anomalies.

Since these tunneling anomalies are associated with the 
charge propagation in the conductor, the exponent depends on the 
junction geometry. If tunneling occurs into the end of a tube, 
the electron can propagate away from the tunneling site only 
in one direction. In contrast, for bulk tunneling the net 
propagation is faster since two different directions are available. 
This implies that the exponent for bulk tunneling is reduced roughly 
by a factor 2 when compared to end tunneling. This is seen 
immediately for the classical resistor case. $R$ is divided by 
two since it is the sum of two resistances in parallel. 
In case of ballistic resistor $\alpha^{end} \sim 2\alpha^{bulk}$ 
in the limit of a small $g$ (large interaction strength). 

Our measurements qualitatively agree with expectations 
for tunneling anomalies associated with 
charge propagation in the conductor. The tunneling 
conductance displays power law behavior both in temperature 
and bias voltage. The observed exponents are geometry dependent 
and obey $\alpha^{end} \approx 2\alpha^{bulk}$.  However, neither 
is in quantitative agreement with our measurements, as we now discuss.

We will first focus on the classical series resistor model. In Fig.~2 the 
resistance between the two outer electrode is $R_{0}$=4.7k$\Omega$ at 2K. 
The tunnel barrier is attached to two resitances $R_{0}/2$ in parallel. 
This corresponds to $\alpha=R_{0}/4(h/2e^{2})=0.09$. This 
value is not in agreement with the experimental value of 0.36. 
Furthermore, we have studied many other samples with different 
lengths, and therefore different resistances, but obtain roughly 
the same exponent. We therefore observe no clear relationship 
between the exponent and the resistance of the MWNT.

This oversimplified model can be improved if instead of a 
bare resistor a RC transmission line is taken\cite{nazarov}. C is the 
capacitance between the tube and the backgate lying 1 micron beneath. 
However, comparing our measurements with the equations given in ref [21] 
gives also bad agreement. The voltage dependence of 
the conductance deviates from a power law and the suppression 
of the conductance is much narrower in energy. 

We now compare our results to the LL model. It has been successively 
used to interpret tunneling in SWNTs\cite{bockrath,yao,postma}, which 
are well-suited to this model because of the 
large elastic length and the fact that only two subbands contribute 
to transport. MWNTs, on the other hand, have 
shorter elastic length and have a large number of conducting 
channels\cite{kruger}. Theoretical calculations\cite{matveev} have 
shown that the exponent is reduced by $N^{-1/2}$ for a wire with $N$ 
channels. The exponent is expected to be 
further suppressed because the inner cylinders participate 
in screening, leading to suppression\cite{egger2} proportional to 
$M^{-1/2}$ where $M$ is the number of inner metal cylinders. 

MWNTs with 10nm diameter have typically $\sim$5 
inner metal cylinders. Recent experiments on the 
electrochemical doping of MWNTs\cite{kruger} demonstrate 
that the number of occupied subbands is large, on the 
order of $\sim$10. The relative high number of modes 
together with the inner cylinders is expected to 
decrease the exponent for MWNTs substantially when 
comparing with SWNTs. The suppression factor is estimated to be 
$\sim$3. This is not what we observe. Indeed, 
the exponents $\sim$0.3 and $\sim$0.6 for bulk and end tunneling 
in MWNTs are similar to those observed for SWNTs. 
The exponents for SWNTs are ranging between 0.26 and 0.38 
for bulk tunneling, and between 0.5 and 1.1 for end tunneling. 
Both of these are remarkably close to the MWNT values. 
Whether this is merely a coincidence or reflects a unified 
underlying physical origin is not clear.

In conclusion, we have shown that tunneling conductance in 
MWNTs vanishes at low energies in a manner well-described by a 
power law in the energy of the tunneling electron. This behavior 
is most likely caused by suppressed charge propagation of the 
injected quasi particle away from the tunneling site. However, 
neither a model of the MWNT as a diffusive conductor nor as a 
multichannel ballistic conductor gives a satisfactory account for 
our results. Future work, both experimental and theoretical, 
is necessary to clarify the nature of the tunneling suppression that we observe.
 
We acknowledge D.~Averin, R.~Egger, E.~Graugnard, M.~Fuhrer, 
P.~Kim, Y.~Nazarov and S.~Tans for discussions. 
We thank J.-P.~Salvetat, L.~Forro, A.~Rinzler and R.~Smalley 
for the nanotube materials. This work was supported by DOE, by 
DARPA and by the Swiss NSF.

$^{*}$ present address: TUDelft, 2628CJ Delft, Netherlands.

\newpage

\begin{figure}
\includegraphics[width=8.1cm]{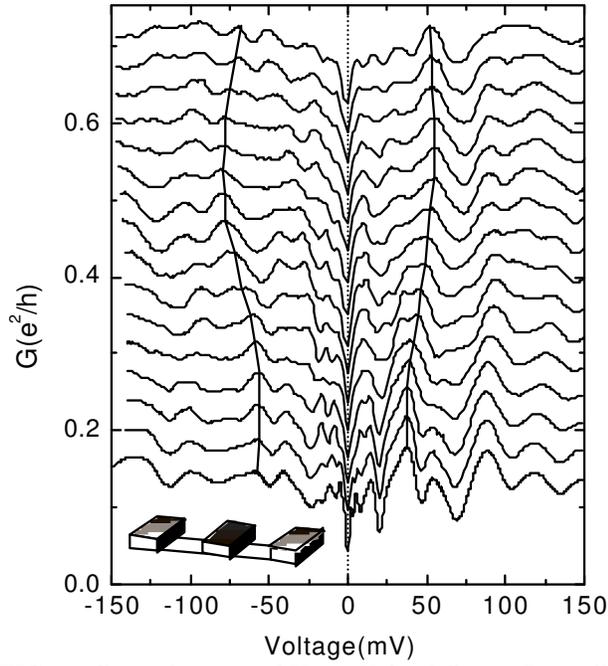}
\caption{$G$ as a function of $V$ at 2K for different $B$
parallel to the tube, 
$B$=0, 1,$\cdot \cdot \cdot$15T. Curves are offset for clarity. The displacement of 
two well discernable peaks are indicated by lines as guide to the eyes. 
In inset a schematic of the MWNT attached to 3 electrodes separated by 350nm. 
The inner electrode has a high resistance contact to the tube.}
\end{figure}

\begin{figure}
\includegraphics[width=8.1cm]{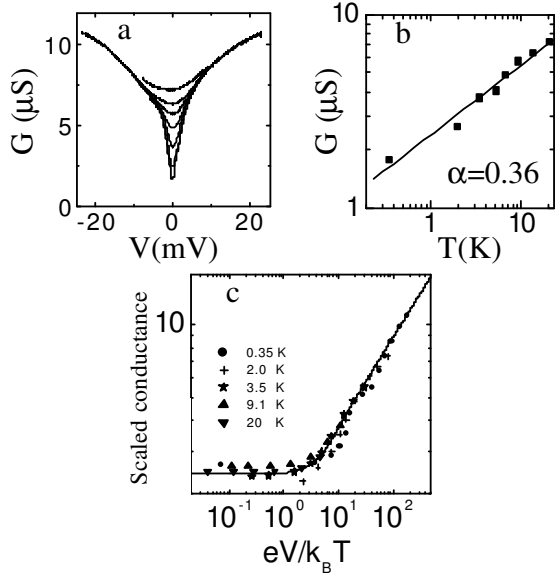}
\caption{(a) $G$ of a second MWNT as a function of $$V for different $T$, from 0.35K to 20K. 
(b) $G$ at zero voltage plotted as a function of $T$ in a double logarithmic scale. 
(c) $G \cdot T^{-\alpha}$ versus $eV/k_{B}T$. At large voltage the slope of the line is 0.36.}
\end{figure}

\begin{figure}
\includegraphics[width=8.1cm]{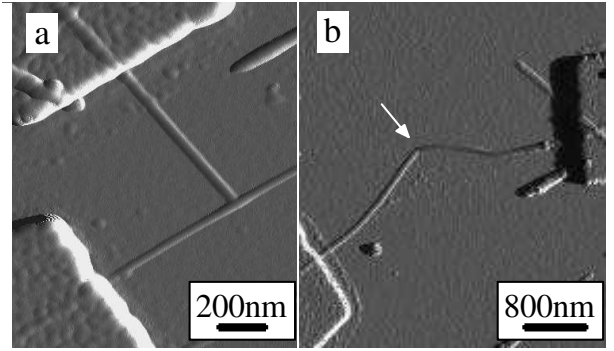}
\caption{(a) End-bulk junction. (b) End-end junction. 
The arrow indicates the position of the junction.}
\end{figure}

\begin{figure}
\includegraphics[width=8.1cm]{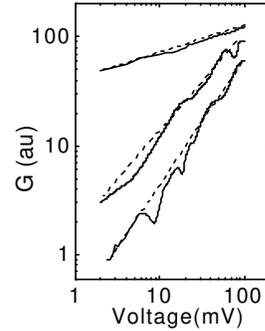}
\caption{$G$ as a function of $V$ in a double logarithmic scale for 
a Au-MWNT, an end-bulk and an end-end junction. The slopes of 
the corresponding lines are 0.25, 0.9 and 1.24. }
\end{figure}


\begin{references}
\bibitem{note1} Measurements on other samples containing a backgate 
show similar peaks, 
which are mostly independent on gate voltage. This allows to rule out 
an origin due to coulomb blockade or to interference.
\bibitem{note2} This Aharanov-Bohm effect is different than the 
one observed in ref [9], which is based on interference effect.
\end{references}
\end{document}